\documentstyle[12pt]{article}

0

\begin{document}

\hspace{12cm} hep-th/9604057

\begin{center}

{}~\vfill

{\large \bf  Integrability, Jacobians and Calabi-Yau 
threefolds\footnote{Based on a lecture given by C. Gomez at the ``VIII
Regional Meeting on Mathematical Physics'', October 1995, Iran.}}

\end{center}

\vspace{10 mm}

\begin{center}

{\bf C\'{e}sar G\'{o}mez$^a$, Rafael Hern\'{a}ndez$^{a,b}$ and 
Esperanza L\'{o}pez$^c$} 
  
\vspace{7 mm}
  
$^b${\em Departamento de F\'{\i}sica Te\'{o}rica, C-XI,
Universidad Aut\'{o}noma de Madrid, \protect \\
Cantoblanco, 28049 Madrid, Spain}
 
\vspace{5mm}  
  
$^a${\em Instituto de Matem\'{a}ticas y F\'{\i}sica Fundamental,
CSIC, \protect \\ Serrano 123, 28006 Madrid, Spain}  
 
\vspace{5 mm}
  
$^c${\em Institute for Theoretical Physics, University of
California, \protect \\ Santa Barbara, CA 93106-4030}

\end{center}      

\vspace{15mm}
   
\begin{abstract}
The integrable systems associated with Seiberg-Witten geometry 
are considered both from the Hitchin-Donagi-Witten gauge model 
and in terms of intermediate Jacobians of Calabi-Yau threefolds. 
Dual pairs and enhancement of gauge symmetries are discussed on 
the basis of a map from the Donagi-Witten ``moduli'' into the 
moduli of complex structures of the Calabi-Yau threefold.
\end{abstract}

\pagebreak

%%%%%%%%%%%%%%%%%%%%%%%%%%%%%%%%%%%%%%%%%%%%%%%%%%%%%%%%%
%%%%%%%%%%%%%%%%%%%%%%%%%%%%%%%%%%%%%%%%%%%%%%%%%%%%%%%%%

Looking for a unified point of view to describe the great amount 
of exciting new results appearing in the last few months in the 
context of string theory can be considered at this moment premature 
and maybe a pure question of taste. Nevertheless, in this lecture we 
would like to present a conjectural unified approach supported by 
some evidence \cite{U} and based on the mathematical 
structure of integrable systems \cite{R,DW,DM}.

\section{Integrability and Seiberg-Witten geometry}

A very elegant way to describe Seiberg-Witten geometry \cite{SW,SW2} 
for $N\!=\!2$ non abelian gauge theories, is \cite{DW} by means of families 
of complex tori \cite{DM}.

Let
\begin{equation}
\pi : X \rightarrow U
\label{1}
\end{equation}
with fiber $X_u$ ($u \in U$) a complex tori of dimension $g$. We will 
assume that$X$ is a symplectic manifold possessing a closed non 
degenerate holomorphic two form $w$ and that dim$\hspace{2mm} U$ 
is equal to $g$. Given a basis $\gamma_i (u)$ $i=1,..,g$ of 
$H_1 (X_u,Z)$, we define ``special coordinates''
\begin{equation}
a_i(u) = \oint_{\gamma_i(u)} \lambda,
\label{2}
\end{equation}
where
\begin{equation}
w= d \lambda.
\end{equation}

These special coordinates define hamiltonian vector fields which are 
tangent to $X_u$. Denoting $H_g$ the Siegel upper half space of 
$g\times g$ symmetric complex matrices $Z$ with $\mbox{Im} 
\hspace{2mm} Z >0$, the period map
\begin{equation}
p: U \rightarrow H_g 
\label{4}
\end{equation}
is defined in such a way that the complex tori $X_u$ is given by 
$C^g/ \Lambda_{p(u)}$ where $\Lambda_{p(u)}$ is the lattice generated 
by $(1, Z_u=p(u))$. The ``dual'' set of special coordinates is 
defined by
\begin{equation}
a_{i}^D (u) = \oint_{\gamma_{i}^D (u)} \lambda,
\end{equation}
and the ``duality'' relations between $a_i$ and $a_{i}^D$ are 
determined by the period map (\ref{4}) as follows \cite{DM}
\begin{equation}
d a_i(u) = p_{ij}(u) d a_{j}^D (u),
\label{6}
\end{equation}
which for $g=1$ gives the well known duality relation \cite{SW} for 
$SU(2)$.

The physical interpretation of the family (\ref{1}) consist in 
associating $U$ to the moduli space determined by some flat potential 
of a $N\!=\!2$ gauge theory, and the fiber $X_u$, together with equations 
(\ref{2}) and (\ref{6}), with the BPS mass spectrum formula.

\vspace{5mm}

\section{Hitchin's gauge models and Prym varieties}

There exits a way to associate with a given gauge group $G$ and a 
Riemann surface $\Sigma$, a family (\ref{1}) of complex tori 
\cite{H,DM,DW}. This procedure is based on the concept of spectral 
curve.

Let us consider as $\Sigma$ the elliptic curve $E_{\tau}$
\begin{equation}  
y^2=(x-e_{1}(\tau))(x-e_{2}(\tau))(x-e_{3}(\tau)),
\end{equation}
and let $\phi$ be a Higgs field defined on $E_{\tau}$ as a 
holomorphic 1-form valued in the adjoint representation of $G$. 
The spectral curve $C$ is defined by
\begin{equation}
\mbox{det} (t-\phi) =0.
\label{8}
\end{equation}

Denoting $J(C)$ the Jacobian of $C$, which for $G=SU(2)$ will have 
dimension two, we can define the Prym-Tjurin subvariety 
$P(C,\sigma)$ as
\begin{equation}
P(C,\sigma)=(1-\sigma) J(C),
\end{equation}
where $\sigma$ is the automorphism
\begin{equation}
\sigma: t \rightarrow -t
\end{equation}
of the curve (\ref{8}) for $\phi$ valued in the adjoint of $SU(2)$.

Let us now consider the Higgs invariant polynomial $tr \phi^2$, 
which is a quadratic differential on $E_{\tau}$. We can evaluate 
$tr \phi^2$ on the basis (one dimensional) of holomorphic two 
differentials of 
$E_{\tau}$. Let us denote by $u$ the corresponding coefficient. The 
family (\ref{1}) of complex tori for $SU(2)$ is now defined 
\cite{DW} with $U$ parameterizing the different values of 
$tr \phi^2$, and with the fiber $X_u$ given by the Prym-Tjurin 
variety $P(C_u,\sigma)$.

It is important here to stress the difference between 
Hitchin-Donagi-Witten construction and the way to associate spectral 
curves to Lax-hamiltonians systems \cite{PM}. In this second case 
one starts with a map $A: P^1 \rightarrow {\cal G}$ on some Lie 
algebra $\cal G$. The spectral curve $C$ defined by
\begin{equation}
\mbox{det}(z \hspace{1mm} {\bf 1} - A(\xi))=0
\label{11}
\end{equation}
is a sheeted covering $p:C \rightarrow P^1$ with fiber at a point 
$\xi_0$ being identified with the set of extremal weights $\{ 
\lambda_1,..,\lambda_r\}$ of $h(\xi_0)$ (the centralizer of 
$A(\xi_0)$). The Weyl group is acting by permutations $S_d$ on this 
set of weights (see reference \cite{PM} for more technical 
details). In the Hitchin-Donagi-Witten approach one starts on a 
general reference Riemann surface, reducing for the cases of 
physical interest to $E_{\tau}$. The equivalent to $A(\xi)$ in 
(\ref{11}) is the Higgs field defined on $E_{\tau}$, and the way 
we parameterize this Higgs field is by the coordinates $u_i$ of the 
invariant polynomials relative to the basis of the corresponding 
$q$-holomorphic differentials of the reference surface $E_{\tau}$. 

The interest of working on $E_{\tau}$ is that we can define an 
alternative family of complex tori keeping fixed the Higgs field 
while changing the moduli $\tau$ of the reference Riemann surface 
$E_{\tau}$. Notice that working on $E_{\tau}$ is not affecting 
the way the Weyl group is acting on the spectral sets.

\vspace{5mm}

\section{Intermediate Jacobians for threefolds and dual pairs}

In the previous section we have discussed a model for families of 
abelian varieties based on Hitchin's construction. The ingredients 
were a reference Riemann surface $\Sigma$ and a Higgs system 
defined on $\Sigma$ with gauge group $G$. A different geometrical 
model for families of complex tori can be obtained for Calabi-Yau 
threefolds using Griffiths intermediate Jacobians \cite{G}. Given 
a Calabi-Yau threefold $Y$, let us denote by $J(Y)$ the 
corresponding intermediate Jacobian (see reference \cite{DM} for 
definitions). The dimension of $J(Y)$ is given by $h_{2,1}(Y)$ 
and the family of complex tori is obtained by fibering $J(Y)$ on 
the moduli of complex structures of $Y$. By means of the 
intermediate Jacobian $J(Y)$ we map 3-cycles of $Y$ into 1-cycles 
of $J(Y)$. Recent results in dual pairs \cite{DP,KV} strongly 
motivate us to look for a way to connect the Hitchin model based 
on the data $(E_{\tau},G)$ and Griffiths 
construction based on a Calabi-Yau threefold $Y$, namely 
interpreting these two constructions as a sort of dual pair with
\begin{equation}
\begin{array}{rcl}
\tau & \rightarrow & \mbox{dilaton} \\
\mbox{rank} \hspace{2mm} G & \rightarrow & h_{2,1}(Y)-1 \\
\mbox{1-cycles of} \hspace{2mm} J(C) & \rightarrow & 
\mbox{3-cycles of} \hspace{2mm} Y. 
\end{array}
\label{12}
\end{equation}
 
Relations (\ref{12}) should be interpreted at this point only as 
a conjectural framework. Before going into a more detailed 
discussion let us make some general comments.
\begin{itemize}
\item[i)] Intermediate Jacobians were used in \cite{DM} 
(see also \cite{W}) to define integrable systems. In order to do 
that it is important to enlarge the Jacobian by adding the 
holomorphic gauge. Namely, in physics terminology, by including 
the graviphoton. We will not discuss this important issue in this 
note. 
\item[ii)] Taking into account that we are inspired by the 
heterotic-type II dual pairs, we should consider Calabi-Yau 
threefolds that are $K_3$-fibrations \cite{KLM}. For these 
fibrations we know \cite{AL} that the size of the $P^1$-basis is 
connected with the heterotic dilaton. Following (\ref{12}) we 
should try to connect this size with the value of $\tau$ for the 
reference surface in the Hitchin-Donagi-Witten construction.
\item[iii)] In the Donagi-Witten approach we have two types of 
1-cycles, namely the ones in $J(C)$ coming from the Jacobian of 
the surface $E_{\tau}$ and the rest. In the next section we will 
review some of the results of \cite{U} proposing a physical 
interpretation for both types of one cycles. 
\end{itemize}

\vspace{5mm}

\section{From Donagi-Witten moduli to a threefold moduli}

We will consider the case of $G=SU(2)$ and $E_{\tau}$ as the 
reference Riemann surface. We parameterize the corresponding 
spectral curve by $(\hat{u},\tau)$, where $\hat{u}\equiv u/
(\frac{1}{2}m^2)$ and with $m$ fixing the residue of the pole of 
the Higgs field on $E_{\tau}$ at infinity (see \cite{DW}). 
Depending on what type of 1-cycle degenerates we can 
differentiate the following two set of singularity 
loci\footnote{The loci grouped as type 2 imply the
contraction of all 1-cycles in $J(C)$ for $SU(2)$, both 
coming from the reference surface $E_{\tau}$ and the rest.} 
\begin{equation}
\mbox{type 1:} \left\{ 
        \begin{array}{lcl}
        \hat{{\cal C}}_{0}  & \equiv & \{ \hat{u}(\tau)=
        \frac {3}{2}e_{1}(\tau) \} \\
        \hat{{\cal C}}_{C}^{(1)} & \equiv 
        & \{ \hat{u}(\tau)=e_{3}(\tau) + \frac {1}{2}e_{1}
        (\tau) \}, \\
        \hat{{\cal C}}_{C}^{(2)} & \equiv & \{ \hat{u}(\tau)=
        e_{2}(\tau) + \frac {1}{2}e_{1}(\tau) \} 
        \end{array}   \right.
\mbox{type 2:}  \left\{ 
        \begin{array}{lcl}
        \hat{{\cal C}}_{\infty}  & \equiv & \{ \tau= i \infty 
        \: / \: \epsilon=0  \} \\       
        \hat{{\cal C}}_{1}^+     & \equiv & \{\tau = 0 \: / \: 
        \epsilon=1  \} \\
        \hat{{\cal C}}_{1}^-     & \equiv & \{\tau = 1 \: / \: 
        \epsilon=-1 \},
        \end{array}   \right.
\label{13}
\end{equation}
with $\epsilon=e^{i \pi \tau}$.

By blowing up the crossing between $\hat{{\cal C}}_{C}^{(1)}$, 
$\hat{{\cal C}}_{C}^{(2)}$ and $\hat{{\cal C}}_{\infty}$ we get 
an exceptional divisor $E$ parameterized by $\tilde{u} \equiv 
\hat{u}/ \epsilon$, that we can identify with the $N\!=\!2$ 
Seiberg-Witten quantum moduli for $SU(2)$.

The action of $T:\tau \rightarrow \tau +1$ on $(\tilde{u},
\epsilon)$-variables is given by
\begin{eqnarray}
\tilde{u} & \rightarrow & - \tilde{u}, \nonumber \\
\epsilon  & \rightarrow & - \epsilon.
\label{14}
\end{eqnarray}
If we quotient by the action (\ref{14}) we get the 
$F_2$-Hirzebruch space
\begin{equation}
\xi \zeta = \eta^2,
\end{equation}
with 
\begin{equation}
\xi \equiv \tilde{u}^2,\: \: \: \: \: \: \eta \equiv 
\frac {\tilde{u}}{\epsilon},
\: \: \: \: \: \: \zeta \equiv \frac{1}{\epsilon^2}.
\end{equation}
This is the space we want to relate with the moduli of complex 
structures of the mirror of $WP_{11226}^{12}$ \cite{C,Y}, 
with $h_{2,1}=2$, and defining polynomial
\begin{equation}
p=z_{1}^{12}+z_{2}^{12}+z_{3}^{6}+z_{4}^{6}+z_{5}^{2}-12 \psi 
z_{1}z_{2}z_{3}z_{4}z_{5}- 2 \phi z_{1}^{6}z_{2}^{6}.
\end{equation} 
Using variables $x=\frac{1}{\phi^2}$ and 
$y=\frac{-\phi}{864 \psi^6}$, 
the blow up of the tangency between the conifold locus and the 
weak coupling locus $\{y=0\}$ produces an exceptional divisor 
parameterized by $\frac{x y^2}{(1-x)^2}$. Identifying the loci 
$\hat{{\cal C}}_{C}^{(1)}$ and $\hat{{\cal C}}_{C}^{(2)}$ with 
the two branches of the conifold, $\hat{{\cal C}}_{\infty}$ 
with the weak coupling locus $\{y=0\}$, 
and identifying the blow ups in both spaces (the $F_2$-space 
$(\hat{u},\tau)$ and the Calabi-Yau moduli), we get
\begin{equation}
\frac{y x^2}{(1-x)^2}=\frac{\epsilon^2}{\hat{u}^2}=
\frac{1}{\tilde{u}^2}.
\label{17}
\end{equation}
This is the point particle limit described in \cite{KKLMV}, in 
which the string moduli exactly reproduces the quantum moduli 
space for $N\!=\!2$ $SU(2)$ Yang-Mills.
Notice also that (\ref{17}) goes into the direction (\ref{12}) 
of interpreting the moduli $\tau$ of the reference Riemann 
surface $E_{\tau}$ as the dilaton. By this identification the 
contracting 3-cycles associated with conifold like 
singularities \cite{S} are mapped into 1-cycles singularities 
of the type 1 (\ref{13}).

The very symmetric structure of the loci 
$\hat{{\cal C}}_{C}^{(1)}$, $\hat{{\cal C}}_{C}^{(2)}$ and 
$\hat{{\cal C}}_{1}^+$, $\hat{{\cal C}}_{1}^-$ in (\ref{13}), 
motivates us to try to extend the relation between 
Donagi-Witten moduli and the threefold moduli beyond the point 
particle limit.

In order to analyze the singularities of type 2, we should 
work out the family of abelian varieties obtained by changing 
the moduli of $E_{\tau}$ while maintaining fixed the 
(4-dimensional) Higgs field, i.e. the value of $u$. Recall 
that $u$ was identified with the coefficient that arises when 
we expand the invariant polynomial $tr \phi^2$ in the basis 
of quadratic differentials of $E_{\tau}$, where $\phi$ is the 
Higgs field on $E_{\tau}$. Denoting by $C_{\tau,u}$ the 
corresponding spectral curve defined by (\ref{8}), then we can, 
for fixed $u$, define a family of abelian varieties over the 
moduli of $E_{\tau}$ with fiber the Prym manifold 
$P(C_{\tau,u})$. At the singular loci of type 2 the 1-cycle of 
$E_{\tau}$ contracts to a point. For this family of abelian 
varieties the $\tau$-plane is playing the role of moduli space 
and the $Z_2$ centralizer of $Sl(2;Z)$ the one of the classical 
Weyl group \cite{U3}.  
By the identification of the $\tau$ and the heterotic dilaton 
$S$, and the relation between $S$ and the $y$ variable
\cite{KV}, it would be natural to map the loci 
$\hat{{\cal C}}_{1}^+$, $\hat{{\cal C}}_{1}^-$ with the 
two branches $\{\sqrt{y}=\pm1\}$ ($\{\phi=\pm1 \}$) of the 
strong coupling locus $\{ y=1 \}$. Moreover, the locus 
$\{ y=1 \}$ corresponds to a wall of the 
K\"ahler cone in the moduli space of the mirror Calabi-Yau, 
which can be interpreted as a topology changing singularity
with non perturbative enhancement of non-abelian gauge 
symmetry \cite{ES,BSV,KM}.
In reference \cite{ES} this enhancement of symmetry was 
determined by the type of ALE space describing the singular 
Calabi-Yau manifold, being $G'=SU(2)$ for $WP_{11226}^{12}$. 
For the family of abelian varieties over 
the $\tau$-plane we have described above, the monodromies 
around the singular loci of type 2 are, up to classical 
Weyl monodromies in the centralizer of $Sl(2;Z)$, conjugated 
to $T^2$, in preliminary agreement with the results of 
\cite{ES}\footnote{In this general discussion we can only
compare monodromies that in \cite{ES} depend on the genus
$g$ of the curve of singularities, which 
determines the number of hypermultiplets in the adjoint 
of $G'$ that also acquire zero mass at the enhancement of 
symmetry locus.}.

A temptative map
\begin{equation}
{\displaystyle x=\frac {3/2e_{1}(\tau)}{3/2e_{1}(\tau)-
\hat{u}},}
\hspace{1cm} {\displaystyle \sqrt{y}=-
\frac {e_{2}(\tau)-e_{3}(\tau)}{3e_{1} (\tau)},}
\label{18}
\end{equation}   
between the $(\hat{u},\tau)$-plane and the Calabi-Yau threefold 
moduli was proposed in \cite{U}. By this map, the 
correspondence between the conifold locus and 
$\hat{{\cal C}}_{C}^{(1)}$, $\hat{{\cal C}}_{C}^{(2)}$, and 
$\{y=1\}$ and $\hat{{\cal C}}_{1}^+$, $\hat{{\cal C}}_{1}^-$ 
described above are satisfied \cite{U2}. The map (\ref{18}) 
reproduces the correct point particle limit of \cite{KKLMV}. 

The pull back by this map of the complex tori defined by the 
intermediate Jacobian of the Calabi-Yau should map the 3-cycles
in $H_3(Y)$ related with conifold singularities to 1-cycles of 
$J(C)$ which are not coming from $E_{\tau}$, and the 3-cycles 
in $H_3(Y)$ related to the non-perturbative enhancement of 
non-abelian gauge symmetries to the part of the $J(C)$ coming 
from the reference surface $E_{\tau}$
\begin{eqnarray}
\pi: J(C) & \rightarrow & J(E_{\tau}) \nonumber \\
\mbox{Ker} \hspace{2mm} \pi & \leftrightarrow & 
\mbox{enhancement of symmetry 3-cycles}
\end{eqnarray}

To finish let us just put the question on the possible 
(F-theoretical) physical meaning of models defined on 
reference Riemann surfaces with genus bigger than one.

\section{Final Remarks}

There are two things we want to stress. The first one is 
connected with duality. To keep alive the $\tau$ parameter, as 
it is natural to do in the Donagi-Witten framework, allows to 
define the action of the $Sl(2;Z)$ duality transformations in 
a very natural way as the modular transformations on $\tau$, 
and to induce them by the map (\ref{18}) into the Calabi-Yau 
moduli.

The second is related with the crucial role of integrability. 
The abstract structure of a moduli manifold fibered by complex 
tori defined by the lattice of BPS-particles, admits two 
interesting mathematical models. The one of Hitchin related to 
gauge theories on a Riemann surface, and the one defined using 
intermediate Jacobians related to Calabi-Yau threefolds. It 
would be appealing if that general framework could help in a 
unified understanding of dual pairs, string duality and non 
perturbative enhancements of gauge symmetry. The previous 
lecture was a modest attempt of that.  

\vspace{5mm}

{\bf Note:} As we were finishing this note it appeared the 
paper \cite{KLMVW}, where the question of integrability and 
Calabi-Yau $K_3$-fibrations is considered.

\vspace{20 mm}
  
\begin{center}
{\bf Acknowledgments}
\end{center}

E.L. would like to thank K. Landsteiner for useful consversations.
This work is partially supported by European Community grant 
ERBCHRXCT920069 and by grant PB92-1092. The work of R. H.
is supported by U.A.M. fellowship. The work of E. L. is supported
by C.A.P.V. fellowship.

\end{document}